\documentclass[twocolumn]{svjour3}         
\smartqed  
\usepackage{graphicx}
\begin{document}

\title{Cognition and Emotion: Perspectives of
       a Closing Gap
}


\titlerunning{Cognitions and Emotion}   

\author{Claudius Gros  
}


\institute{Claudius Gros \at
         Institute of Theoretical Physics\\
         J.W. Goethe University Frankfurt\\
         60054 Frankfurt/Main, Germany}

\date{Received: date / Accepted: date}

\maketitle

\begin{abstract}
The primary tasks of a cognitive system is to
survive and to maximize a life-long utility
function, like the number of offsprings.
A direct computational maximization of
life-long utility is however not possible
in complex environments, especially in the context,
of real-world time constraints. The central role of
emotions is to serve as an intermediate layer
in the space of policies available to agents and
animals, leading to a large
dimensional reduction of complexity.

We review our current understanding of the
functional role of emotions, stressing
the role of the neuromodulators mediating 
emotions for the diffusive homeostatic control
system of the brain. We discuss a recent proposal, 
that emotional diffusive control is characterized, 
in contrast to neutral diffusive control, by 
interaction effects, viz by interferences between
emotional arousal and reward signaling.
Several proposals for the realization of
synthetic emotions are discussed in this
context, together with key open issues
regarding the interplay between emotional
motivational drives and diffusive control.

\end{abstract}

\section{INTRODUCTION}

The apparent dichotomy between the progression of
human cognitive achievments during the last
millenia and the pervasiveness 
of affective behavioral patterns in everyday life has 
fascinated philosophers from the beginning of times and 
is subject to libraries of literature. Quite often
it is assumed in this context, that the emotional
groundings of human behavior are somehow a leftover
heritage from our more `animal-like' predecessors,
and that `rational behavior' is more appropriate
for humans and `superior' to affective conducts.

This popular appraisal of emotions is however utterly 
wrong, since emotional control has functional purposes 
which are indispensable for full fledged cognitive 
systems and which cannot be substituted by cognitive 
information processing. We review here our current
understanding of the human emotional control system,
from the functional point of view, including 
considerations from dynamical systems theory.
The emphasis will be on general properties, 
being of possible relevance not only for the 
understanding of the brain but also for eventual 
human-level artificial intelligences. 

The emotional control system is, from the evolutionary
perspective, a specialization of homeostatic control.
Myriads of auto-regulative processes are maintaining our
bodily functions at every moment and we are conscious of
only a small subset having an emotional context. The
question then arises which traits are characteristic for
automatic and which for emotional homeostatic processes. In this
context we discuss a recent proposal suggesting that
emotional processes have a genetically determined preferred
level of activation and are characterized by interaction effects
within the homeostatic control system, whereas neural
control processes are void of any genetically preferred 
levels of activation. 

\subsection{Functional emotions}
\label{subsect_functional_emotions}

We are around nowadays only because our
ancestors managed to survive and to produce
offsprings, the basic prerequisites for 
evolutionary fitness. Daily survival 
can be regarded as a homeostatic process
for keeping the basic bodily parameters, 
via interaction with the environment,
in their proper range. Being a product
of evolutionary selection processes, 
the emotional constituents of our self
must necessarily contribute 
to survivability \cite{rolls05}.\footnote{Subject
to evolutionary pressure are all traits which
influence Darwinian fitness. An emotional
arousal typically leads to behavioral consequences
and behavior is a primary toehold for
evolutionary selection. This evolutionary
perspective sometimes contrast with our
daily experiences, as certain human emotions 
are routinely viewed to be more a
handicap, instead of being benficial, when
living in modern societies.}
One can therefore classify the overall 
functionality of our emotional control 
system as homeostatic 
\cite{damasio94,damasio99}.\footnote{One 
may actually wonder, against the background 
of the fact that our emotional suit
contributes in important ways to our Darwinian fitness, 
why then are emotions portrayed often as irrational and 
counterproductive in everyday life.}

The perspective of this short review is the functional
role of emotions. Instead of the plain expression
``emotions''we will be using mainly the terminology
{\em ``emotional control system''}, which reflects
more precisely the functional role of emotions as 
part of the homeostatic control system of the brain.
Emotions are part of the web of physical and biochemical 
processes occurring which, as a whole
is denoted {\em ``cognitive system''} \cite{grosBook08}.
The cognitive system has the task to keep its
support unit (the body) alive, as well as its
wetware (the brain). A cognitive system
is what has been called `organismic' in the
framework of enactive artificial intelligences
\cite{froese09,ziemke01}. The term cognitive system places
then emphasis on the dynamical system perspective; the
physical brain tissue is not identical with human 
consciousness and affection, but the collection of 
interacting neural, physical and chemical dynamical 
processes occurring at every moment of time.

\subsection{Natural and synthetic emotions}
\label{subsect_natural_synthetic_emotions}

An extensive literature is devoted to the question
of the introperspective content of emotions as
they are experienced, see \cite{barrett07}.
The spectrum of emotional experiences is
vast, ranging from plain fear conditioning
\cite{ledoux00} and romantic love \cite{aron05},
to the complexity of social interactions 
\cite{blakemore04,lieberman07}. Emotional
expressions play a paramount role in social
interactions \cite{adolphs03}, and are
studied increasingly in the context of
human-robot interactions \cite{breazeal03}.
These important issues are not the subject of this
review, they are however closely related to
the core questions regarding the defining functional
characteristics of emotional processes.

It is clearly possible to build humanoid robots 
showing facial expressions, which an anthropomizing
observer would interpret as emotional \cite{duffy03,fong03}.
Emotional expressions by socializing robots may be 
helpful for human-robot interactions but clearly 
do not correspond to true `synthetic emotions',
which need to be related to behavioral control 
\cite{arbib04,ziemke09,vallverdu09}, the focus of the
present review.

\subsection{Homeostasis and diffusive control}

An essential aspect of the living condition is homeostasis,
the active regulation of biological relevant parameters, like
the blood-sugar level or the heart beating frequency. Homeostatic
regulation is also necessary for the internal parameters of 
individual neurons, like membrane conductivities or firing 
thresholds, as well as for networks of neurons \cite{marder06}.
The availability of the neurotransmitter glutamate and
GABA, to give an example, has to be regulated through a 
homeostatic cycle involving the astrocytes \cite{bak06} .

Complex dynamical systems, like the neural net in the brain,
need to retain their overall dynamical properties in a suitable
range, they need to adjust their working point \cite{grosBook08}.
The occurrence of epileptic seizures is an example of what can
happen when homeostasis breaks down. In addition to the
homeostatic regulation, necessary to retain operationability,
neural circuits in the brain also need to allow for transient
modulatory adaption \cite{giocomo07}. This second type of
regulation allows the neural circuits to work in several
different regimes, increasing e.g.\ the relative importance
of afferent input \cite{giocomo07} or the relative importance
of inter-neural competition \cite{krichmar08}.

\begin{figure}[tb]
\centerline{\hfill
\includegraphics[width=0.45\textwidth]{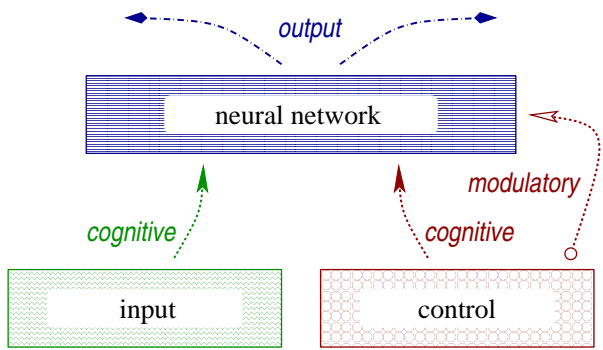}
           \hfill}
\caption{The difference between cognitive and modulatory control.
A neural network is driven by an input layer and its functional behavior
regulated by the control unit. For the same stimulation pattern from the
input layer different outputs are possible, depending on the signals
it receives from the the control unit. When a control 
signal influences the network neurons via direct synaptic 
connections, the control is cognitive; if it affects the parameters 
of the network neurons (dashed line with open head and tail), 
like the firing threshold or the gain, see Eq.~(\ref{eq_modulatory_control}),
the control is modulatory.
        }
\label{figure_control_two}
\end{figure}

Homeostasis is fundamentally diffusive in nature. The need to
regulate the concentration of a given ion or of a certain neurotransmitter
is normally performed non-locally. We are interested here, with
regard to the functional role of emotional control, in the
diffusive control of the properties of neural circuits
by neuromodulators like dopamine or seretonin.

\section{NEUROBIOLOGY OF EMOTIONAL AND DIFFUSIVE CONTROL}

We will now discuss a few selected aspects of the
neurobiology of emotions relevant for understanding
the functionality of emotional control. Our aim is 
however not to provide an extensive review of the 
neurobiological foundations of emotions, for comprehensive 
reviews on the subject see \cite{ledoux00,burgdorf06,pessoa08}.

The neurobiological foundations of emotions are neuromodulators 
like dopamine, seretonin and the ophids \cite{fellous99}.
These neuromodulators are emitted by specialized neurons
originating in quite localized subcortical structures,
like the raphe nucleus and the substantia nigra, 
ascending to cortical areas, in particular to the prefrontal 
cortex, as well as to sub-cortical areas like the amygdala 
and the hippocampus, to 
mention a few exemplary target areas. Emotional experiences,
involving complex recurrent interactions with cognitive 
processing \cite{grey04,ochsner05}, are not identical with 
neuro\-modu\-la\-tor con\-cen\-tra\-tions,
but there is probably no fragile or robust emotional 
experience without the concurrent release of some 
combination of neuromodulators.

A neuromodulator, like dopamine, is
released synaptically. A dopaminergic neuron fires,
like any other neuron, and the depolarization pulse
travelling along its axon activates the dopaminergic
synapses alongside its way. The number of dopaminergic 
neurons is rather small, e.g.\ there are about 7200 
dopaminergic cells in the substantia nigra, each one having 
on average about 370 000 syn\-apses \cite{arbuthnott07}. 
This very large number of syn\-apses per single dopaminergic 
neuron indicates that the release of dopamine produces a
diffusive volume effect. Cells in the target area have 
appropriate receptors on their membrane which will 
change certain membrane properties when activated.

\subsection{Diffusive, modulatory and cognitive control}
\label{subsect_control}

Neuromodulators take their name since they 
modulate the behavior of the neurons in the target
area. Within control theory the term ``modulation'' 
generally just implies that the control effect
is relatively weak with respect to the eigendyamics
of the target, viz modulating and not driving.
Here we will use the term ``modulation'' in a
more restricted sense, with a sharp and qualitative
distinction between modulatory and cognitive
control.

In Fig.~\ref{figure_control_two} we illustrate
the difference between modulatory and cognitive
influence and control. For concreteness let us assume
that a typical neuron in the network layer of
Fig.~\ref{figure_control_two} has a sigmoidal 
activation function,
$$
\sigma(r,\beta,\theta) \ =\ {1\over \mbox{e}^{-(\beta r-\theta)}+1}~,
$$
with an activation threshold $\theta$. The gain $\beta$ encodes
the steepness and $r$ the cognitive input, 
\begin{equation}
r\ =\ \sum_{i\,\in\,\mbox{input}} w_i\, x_i\,+\,
\sum_{j\,\in\,\mbox{control}} v_j\, y_j~,
\label{eq_cognitive_control}
\end{equation}
where the $\{x_i\}$ and the $\{y_i\}$ are neural activity levels
of the input and the control layer respectively and
the $\{w_i\}$ and $\{v_i\}$ the respective synaptic strengths. 
Alternatively, the parameters of the activation function 
$\sigma(r,\beta,\theta)$ of the network-layer neurons
might be modulated by the control layer,
\begin{equation}
\beta\ =\ \beta(y_1,\, y_2,\, \dots),
\qquad\quad
\theta\ =\ \theta(y_1,\, y_2,\, \dots)~.
\label{eq_modulatory_control}
\end{equation}
We can now clarify the different forms of
possible controls.
\begin{itemize}
\item {\bf Cognitive control}\\
      The neurons of the control layer influence
      upstream layers in the same way as any other input, 
      compare Eq.\ (\ref{eq_cognitive_control}), 
      e.g.\ via glutamergic synaptic connections.
\item {\bf Modulatory control}\\
      There are no direct (e.g.\ glutamergic) synaptic
      connections from the control layer to
      upstream layers, viz the $v_j\equiv0$ in
      Eq.\ (\ref{eq_cognitive_control}). The
      influence on upstream layers occurs exclusively
      via the (e.g.\ dopaminergic) modulatory 
      influence on the internal parameter
      of upstream neurons, as in Eq.\ (\ref{eq_modulatory_control}).
\item {\bf Diffusive control}\\
      Diffusive control is modulatory. Modulatory control 
      could still be target specific, as a matter of principle, 
      viz the modulatory effect could be different
      for individual target neurons. Diffusive modulatory
      control is not target specific on the level of
      individual neurons, being a volume effect.
\end{itemize}
The influence of neuromodulators in the brain 
seems to be predominantly diffusive in above
sense \cite{arbuthnott07}. Neuromodulators
can be compared to hormones on a functional
level, which also act diffusively. Hormones are 
synthesized in certain glands, dispensed throughout 
the body via the blood system and act as diffusive 
chemical messengers. Both hormones and neuromodulators
may have lasting as well as phasic effects on their 
respective targets. Stress hormones change the actual
working stage of the body into a fight-or-flight stance,
growth hormones fulfil their task on the other side over 
the course of months and years.

\begin{figure*}[tb]
\centerline{\hfill
\includegraphics[width=0.75\textwidth]{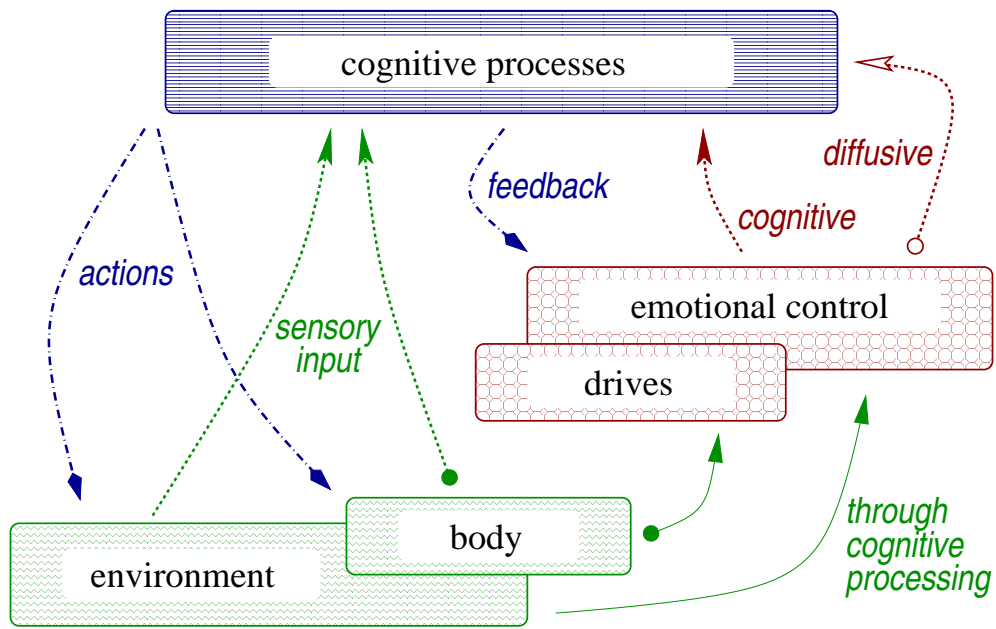}
           \hfill}
\caption{Functional relation between the cognitive processes, 
environment and emotional control. The biological support unit
(the body) and the wetware (the brain) are functionally part 
of the environment. The motivational drives are traditionally
divided into the primary drives and higher emotional control,
the former responsible for securing daily survival, the later
for optimizing life-long Darwinian fitness. The paradigms 
of embodied cognition/emotion state that the proprioceptual
sensory input from the body (lines with filled arrows and circles) 
to the cognitive processes/emotional control
is essential for for cognition/emotions. The paradigm of diffusive
emotional control states that emotions are functionally dependent
on diffusive modulatory control (line with open arrow and circle).
        }
\label{figure_control_emotions}
\end{figure*}

\section{COGNITIVE SYSTEM THEORY}

Let us now consider the overall picture. In
Fig.~\ref{figure_control_emotions} the functional
interdependencies between cognitive pro\-ces\-ses, environment 
and emotional control are illustrated. These 
interdependencies hold both for real-world cognitive 
systems, like the human cognitive system, as well as for 
organismic artificial intelligences. We start the discussion 
by considering the individual components.
\begin{itemize}
\item {\bf Cognitive processes}\\
      All standard conscious and unconscious neural activities, 
      like neural firing and learning via synaptic plasticities,
      belong to the class of cognitive processes. 
\item {\bf Environment}\\
      Notably here is the circumstance, that the support unit
      of the cognitive system, the body, belongs to the 
      environment. It can be acted upon, as any other part of 
      the environment, and the cognitive system can obtain 
      sensory information about it, either via the normal 
      sensory organs or proprioceptionally. The body is 
      however a very special part of the environment
      and plays a central role in the notion of embodiment,
      as discussed further below. An organismic cognitive system
      dies by definition, whenever the support unit ceases to be
      operational.
\item {\bf Emotional control}\\
      The emotional control plays a central role in behavioral control,
      see Sec.\ \ref{sec_netural_emotional}. It is only then distinct
      from cognitive information processing when a diffusive modulatory 
      influence on the cognitive processes is present. Otherwise one may
      subsume the emotional processes under the cognitive processes.
\end{itemize}
Note, that most models for synthetic emotions consider emotional
drives to be state variables interacting with other
state variables, like neural activity levels, cognitively,
e.g.\ via direct synaptic links \cite{vallverdu09,canamero05}.
In these models the emotional states typically have global
effects and are therefore, to a certain extend, special 
state variables. They could be subsumed, nevertheless,
under cognitive information processing,
having analogous functionalities.

\subsection{Cognitive system paradigms}

In Fig.~\ref{figure_control_emotions} the dominant
interactions between the constituent components of
cognitive system and environment are illustrated. 
One needs to remember however that 
environment, emotional control and cognitive processing
have all their own intrinsic and autonomous dynamical 
processes. This state of affairs is self evident for 
the case of the environment, which is generally 
only weakly and locally affected by the actions of the 
cognitive system. But also the cognitive processes 
themselves, viz the neural brain 
activity, have strong and essential
autonomous components \cite{gros05,gros07,gros09}.
The brain is indeed not just a glorified input-output mapper,
driven by the sensory data input stream, but a
self-sustained dynamical system of its own.

There are contrasting views with respect to the 
overall importance of the various interactions 
illustrated in Fig.~\ref{figure_control_emotions},
for a full fledged cognitive system. Disregarding 
many interesting details, these different views
can be stereotyped via their paradigmatic 
assumptions.

\begin{itemize}
\item {\bf Embodied cognition}\\
      The brain receives sensory information both via 
      sensory organs like eye and ears, as well as
      proprioceptual information from the own
      body. The paradigm of embodied cognition states,
      among other things, that the proprioceptual sensory 
      input is not only helpful, but an essential part 
      of cognition \cite{wilson02,anderson03}.
\item {\bf Embodied emotions}\\
      The emotional control system receives information
      both about the environment, preprocessed cognitively,
      as well as direct proprioceptual input from the body. 
      The paradigm of embodied emotions states,
      that the proprioceptual information is an essential
      part of emotions in general and of emotion experience
      in particular \cite{niedenthal07,panksepp05}.
\item {\bf Diffusive emotional control}\\
      The emotional control system might influence the
      cognitive processes both directly  or via
      modulatory processes, as described in
      Sect.~\ref{subsect_control}, with the modulatory
      processes being diffusive volume effects. 
      The paradigm of diffusive emotional control states,
      that a key functionality of natural and synthetic
      emotions involves diffusive control \cite{grosEmo09},
      viz that cognitive emotional control alone may 
      only mimic certain secondary features of 
      emotional control, but not reproduce the core
      functionalities. 
\end{itemize}
The term ``cognitive system'' is widely used in
a range of contexts, sometimes for large-scale
cognitive architectures, more often however for 
specialized algorithms suitable for solving certain well 
defined tasks. Here we use the term cognitive system
as a synonym for ``organismic cognitive system'' 
\cite{grosBook08,froese09,ziemke01},
as corresponding to a full fledged synthetic or
real-world cognitive system, like the human cognitive system.
The above list of paradigmatic assumptions for cognitive
systems is not meant to be exhaustive, many others 
possible architectures have been proposed. We have 
focused here, from the viewpoint of dynamical system
theory, on working principles important for
organismic cognitive systems.

The concept of diffusive emotional control,
which has its groundings in neurobiological
observations, makes a mathematically well 
defined proposal of how to implement synthetic 
emotions. Both emotional control and homeostatic
regulative processes involve modulatory
control via diffusive volume processes.
Indeed, as mentioned in the introduction,
emotional control is evolutionary-wise 
an offspring of homeostatic control.
What is then the difference between
normal or neutral and emotional diffusive
control? This riddle, which is the subject of
Sect.~\ref{sec_netural_emotional}, is not
resolved by the paradigm of diffusive emotional 
control alone.

\subsection{Cognitive feedback}
\label{subsect_cognitive_feedback}

We conclude the overall assessment of an
organismic cognitive system with the
feedback influence of the cognitive 
processes onto the emotional control,
compare Fig.~\ref{figure_control_emotions}.

Cognition and emotions are deeply intertwined
\cite{pessoa08,phelps06} and it is clear that
cognitive processing influences the emotional
control via direct feedback loops. It is possible,
to give an example, to lay relaxed on a couch with
closed eyes, meditating about the happenings of the
last days and experiencing an emotional roller coaster.
Emotions can be triggered both by environmental stimuli
as well as by cognitive processes.

The emotional control system has, in any case, no sensory 
organs of its own. The emotional system cannot obtain direct
information about the environment, all sensory input arriving
to the emotional control is cognitively preprocessed, compare
Fig.~\ref{figure_control_emotions}. It is however
presently a matter of debate exactly to which
cognitive information the emotional control system
has access to. E.g.\ it has been argued \cite{redgrave06},
that the dopamine neurons in the substantia nigra
receive visual information thought the
superior colliculus, which is specialized in
localizing changes of luminance
in the visual field, directing, beside
others, saccadic eye movements. In this
case, the dopaminergic neurons in the substantia nigra
would have access only to a very limited and
specialized sector of cognitive information.

The fundamental drives like hunger and pain are 
less influenced by cognitive feedback, and 
more driven by signaling from the body. The 
distinction between the fundamental drives and 
the higher emotional control can be made precise,
when identifying the drives with the 
``primary survival parameters'' 
\cite{grosEmo09}, viz with the set of parameters
regulating the survival of the body. In this
interpretation the level of blood sugar, the heart
beating frequency and the level of pain, to mention
a few, need to be kept homeostatically in a certain
range by the cognitive system for the support unit to
retain operationability. We die when our blood sugar 
level raises above, or falls below, a certain critical
level.

\section{NEUTRAL VS. EMOTIONAL CONTROL}
\label{sec_netural_emotional}

Our discussion has led us so far to two points:

\begin{itemize}
\item Substantial support from neurobiology indicates,
      that the neuromodulatory system acts via diffusive
      volume processes. The effect of a neuromodulator on
      the neurons in the target area is modulatory and not
      cognitive. The neuromodulators do not affect the 
      actual firing states or the membrane potentials directly, 
      modulating the internal parameters like firing 
      thresholds, learning rates, etc. With emotions being 
      linked to the concurrent release of neuromodulators, 
      these results have important consequence for models 
      of synthetic emotions. Models of emotions involving 
      cognitive emotional control, instead of diffusive control, 
      may be useful for applications, but will not lead 
      in the end to `true synthetic emotions'.

\item The neuromodulatory system is part of the web of
      homeostatic regulations. The vast majority of 
      homeostatic processes occurring in our bodies 
      and in the brain are however neutral, not evoking
      emotional experiences. Which key functionalities
      then differentiate neutral from emotional control?
      This is the question we are addressing in the 
      present section.
    
\end{itemize}
A word of caution. We will propose here a functional 
difference between neutral and emotional homeostatic
control, based both on theoretical as well as on
neurobiological considerations, a proposal which is 
implementable and such also falsifiable. Whether or not 
functional properties of emotional control alone
suffice for an explanation of emotional experiences
is presently however unclear. We start our discussion
by considering the motivational problem.

\subsection{The motivational problem}

Any organismic cognitive system has to take
actions. The question is then which actions and
for which purposes. We propose there that
this motivational problem can be classified, 
cum grano salis, into two separate time windows.
\begin{itemize}
\item {\bf Day-to-day survival}\\
      The task to keep the body healthy on a daily
      basis is job of the primary drives, which take
      their input from survival parameters,
      see Sect.~\ref{subsect_cognitive_feedback} and
      Fig.~\ref{figure_control_emotions}. The survival
      parameters, like the heart beating frequency,
      signal proprioceptually the status of the 
      support unit to the cognitive system. The resulting
      primary drives, like hunger and reproduction, 
      constitute the only set of motivational drives for
      primitive organismic cognitive systems.
      
\item {\bf Life-long Darwinian fitness}\\
      The survival probability, over the lifespan of 
      an individual, and the overall number of 
      offsprings, can benefit from complex behavioral 
      strategies which transcend the requirements for
      day-to-day survival. Optimizing life-long 
      Darwinian fitness is a primary task of emotional 
      control. 
    
\end{itemize}
Two examples: Many animals will engage in social
activities when immediate survival is not at stake,
improving such lifelong survivability (securing the
protection of the kins) and Darwinian fitness
(increasing the chances of obtaining a mate).
Many humans will engage in explorative activities
when being bored, viz when the operational status of
mind and body is close to optimality. General
explorative behaviour, viz exploration without
an explicit goal, may indeed potentially increase
longer-term fitness, e.g.\ when finding a 
shelter for the next winter, or when discovering 
additional nutriment sources for possible 
times of hardness. 

The existence of deep interrelations between 
emotional control, decision making and behavior 
is well established
\cite{dolan02,panksepp03,coricelli07}. Indeed
the notion of empathy, the possibility
to understand deeply the feelings of other persons,
is important in social contexts since empathic 
understanding allows to predict the likely future course
of actions of your counterpart \cite{vignemont06}.
The mere fact that empathy evolved through
evolutionary selection therefore directly implies
that emotional control constitutes a key player 
in behavioral control. 

Emotions have a twofold functionality for behavioral 
control. On one side they establish general 
moods of direct behavioral relevance, like boredom in
the example above. Cognitively evaluated plans and 
targets are, in addition, emotionally weighted, 
a precondition for decision making in the 
framework of both short- and long-term planing.
Policy making is therefore intrinsically dependent,
also for highly developed cognitive systems,
on a solid emotional grounding, the biological
counterpart to the value function used in 
the context of reinforcement learning \cite{sutton98}. 

The here postulated separation of time scales for
instincts and emotional control needs to be 
interpreted non-exclusively. Instincts contribute
to the long-term fitness, but only indirectly 
through a succession of short-term survivals. 
Emotional control can contribute however directly 
both to the prospective life-long fitness and 
to the daily survival rate.

\subsection{Emotional control vs.\ utility maximization}

One may then ask oneself \cite{krichmar08,grosEmo09}, 
which is the advantage of having, in addition to the
primary drives like hunger and reproduction, an
additional layer of motivational drives, the emotional
control. A mainstream hypothesis in artificial
intelligence research assumes, that high-level 
synthetic cognitive systems having life-long utility 
maximization as their sole motivational drive \cite{hutter05},
may eventually be constructed. The artificial cognitive
system would then, in this view, take into consideration 
all facts known about the environment, make a large
computational effort, and maximize directly a 
preset utility function.

Emotional control in mammals works however indirectly.
Emotions do in general not direct the behavior towards 
a straight-forward utility maximization, e.g.\ when
exploring the environment when nothing else is to do.
We believe that general, emotionally induced, 
behavioral strategies are an essential part of
behavioral control in the face of complex environments. 
Any real-world cognitive system is confronted
with a shortage of three types of resources:
\begin{itemize}
\item information about the environment is (extremely)
       limited, 
\item the computational resources (of the brain)
      finite and 
\item the time available for taking decisions 
      is (very) short. 
\end{itemize}
The scarceness of theses three 
resources makes it impossible for an organismic 
cognitive system to optimize directly Darwinian 
fitness over longer time spans, like  weeks, 
months and years. Emotionally regulated behavior 
has proven itself, through selection
and evolution, to be a feasible route to achieve
high levels of Darwinian fitness when only modest
resources (information, computational capabilities and time)
are available.

Emotional control achieves computational effectiveness
by providing general evaluation benchmarks,
making situation-specific evaluations, which
are computationally expensive and very time consuming, 
dispensable. Consider a paleolithic tribe wandering around 
in search for a place to settle. They will likely
direct their pace towards places looking nice,
like a savannah interspersed with patches of forest,
or a small stream with sheltering rocks. This 
emotional valuation, `this place looks welcoming',
is based on a very small subset of potentially
available information. It can be performed very fast
and is computationally affordable, having a substantial 
influence on the longer-term Darwinian fitness of the 
tribe-members.

A utility maximization procedure would correspond
in this situation, on the other side, to a realistic
assessment of the perspectives: how may heads of game
will we be able to hunt here during the next few
months, how many roots and vegetables will we be
able to gather, which is the likelihood of a forest-fire,
and so on. A rigorous utility maximization would clearly
need a substantial investment of time and resources for data 
acquisition and evaluation, a luxury not available
in most situations.

\subsection{Homeostatic interactions}

Emotions correspond to homeostasis at the behavioral level
\cite{craig03}. Is there then a functional difference, which
makes certain homeostatic regulation by neuromodulators 
`emotional', in contrast to automatic homeostatic processes,
which we may term `neutral'? A vast body of clinical data shows 
that emotions and the organization of behavior through 
motivational drives are intrinsically related 
\cite{arbib04,bechara00,naqvi06,coricelli07} 

Highly developed cognitive systems need to acquire
adequate response strategies for given states of 
emotional arousal. When angry, to give an example,
one has to find out which actions are suitable 
for reducing this unpleasant level of arousal.
These response actions are generally not 
genetically predetermined, since quite disparate
environmental situations may lead to 
increased levels of angriness: shooing away 
an annoying fly is an utterly different action 
than settling a quarrel with a spouse.

Response strategies are acquired algorithmically
via reinforcement or temporal-difference learning 
\cite{sutton98}. These learning processes make use
 of reward signals and a given behavioral response 
will be enhanced or suppressed for positive and 
negative reward signals respectively, a well-known 
candidate for a reward signal in the brain being dopamine
\cite{arbuthnott07,doya99}. 
\begin{quote}
Emotional diffusive control is therefore
characterized by a coupling of the regulative event to the 
generation of reward signals for subsequent reinforcement 
learning processes. 
\end{quote}
It has been proposed \cite{grosEmo09},
that this coupling is the characteristic hallmark of
emotional control, differentiating it to neural control
processes.

Which are the conditions for the generation of the reward 
signals coupled to emotional control? Let us come back 
to above example. If we are angry, we will generally try 
to perform actions with the intent of reducing our level 
of arousal. Angriness is reduced when this goal is achieved,
and a positive mood follows, viz a positive reward signal
has been generated, reinforcing the precedent behavior. 
The generation of reward signals is hence coupled to the 
arousal level of the emotional control processes. 
The distinguishing attribute of emotional control is
therefore a genetically predetermined activation or
arousal level. Deviations from this preferred range 
of activation will lead to positive or negative 
reinforcement signals.

We conclude this section with a note of caveat.
The functional characterization of emotional control
given above makes no statement regarding the
preconditions necessary for emotional experiences.
Introspectively the qualia of emotional experiences
is qualitatively different from cognitive reasoning
and thinking. It is presently unclear which attributes
of the emotional circuits are essential for emotional
experiences. Embodiment may play a crucial role in this
respect. Emotions standardly influence the status of 
the body via the release of hormones, like adrenaline,
and we are in part able to sense proprioceptually 
the resulting bodily effects. In this view the experience
of emotions would correspond to proprioceptual 
perception \cite{prinz06}.

\section{CONCLUSIONS}

Discussing the functional role of emotional control
we have identified two core properties:
\begin{itemize}
\item Emotional control via appropriate neuromodulators
      correspond to diffusive modulation of the neural
      activity and not to direct cognitive control.
\item Emotional control is part of the homeostatic
      control system with the difference being that
      the level of emotional arousal is linked to
      the release of positive or negative 
      reinforcement learning signals.
\end{itemize}
These findings, which are based on neurobiological and
clinical observations, have two implications. On one
side they specify mathematically well characterized
functional features of diffusive emotional control 
and are therefore implementable for synthetic 
cognitive systems. They are, on the other hand, 
quite general statements and a myriad of interesting
and important additional details are needed in order 
to obtain a deeper understanding of emotional control.
In this context we have presented an hypothesis
regarding the benefit of emotional control for the
Darwinian fitness of an organismic cognitive system,
arguing that it contributes to the optimization
of both life-long fitness and daily survival, with
the later being the domain of motivational 
control through instincts.

We have emphasized in this review the functional 
differences between emotional control and cognitive 
computation. On the other side we have also pointed out
that both cognition and emotional control are
two indispensable components of any full-fledged
organismic cognitive system. The motivational problem
cannot be solved by cognition alone, any highly developed
cognitive system would remain goal-less stranded and
disoriented without solid and well developed 
emotional groundings.



\addtolength{\textheight}{-0cm}   

\end{document}